# Electron trapping at $SiO_2$/4H-SiC interface probed by transient capacitance measurements and atomic resolution chemical analysis

Patrick Fiorenza[1,*], Ferdinando Iucolano[2], Giuseppe Nicotra[1], Corrado Bongiorno[1], Ioannis Deretzis[1], Antonino La Magna[1], Filippo Giannazzo[1], Mario Saggio[2], Corrado Spinella[1], Fabrizio Roccaforte[1]

[1]*Consiglio Nazionale delle Ricerche – Istituto per la Microelettronica e Microsistemi (CNR-IMM), Strada VIII, n.5 Zona Industriale, 95121 Catania, Italy*
[2]*STMicroelectronics, Stradale Primosole 50, 95121 Catania, Italy*
[*]patrick.fiorenza@imm.cnr.it

## ABSTRACT

Studying the electrical and structural properties of the interface of the gate oxide ($SiO_2$) with silicon carbide (4H-SiC) is a fundamental topic, with important implications for understanding and optimizing the performances of metal-oxide-semiconductor field effect transistor (MOSFETs).

In this paper, near interface oxide traps (*NIOTs*) in lateral 4H-SiC MOSFETs were investigated combining transient gate capacitance measurements (*C-t*) and state of the art scanning transmission electron microscopy in electron energy loss spectroscopy (*STEM-EELS*) with sub-nm resolution. The *C-t* measurements as a function of temperature indicated that the effective NIOTs discharge time is temperature independent and electrons from NIOTs are emitted toward the semiconductor via-tunnelling. The NIOTs discharge time was modelled taking into account also the interface state density in a tunnelling relaxation model and it allowed to locate traps within a tunnelling distance up to 1.3nm from the $SiO_2$/4H-SiC interface. On the other hand, sub- nm resolution STEM-EELS revealed the presence of a *Non-Abrupt* (*NA*) $SiO_2$/4H-SiC interface. The NA interface shows the re-arrangement of the carbon atoms in a sub-stoichiometric $SiO_x$ matrix. A mixed $sp^2/sp^3$ carbon hybridization in the NA interface region suggests that the interfacial carbon atoms have lost their tetrahedral SiC coordination.

**INTRODUCTION**

Today, silicon carbide (4H-SiC) is the most promising wide-bandgap semiconductor material for power electronics applications[1]. In particular, 4H-SiC metal–oxide–semiconductor field-effect transistors (MOSFETs) have been already commercialized and employed as power-switching devices. However, the electrical properties of the $SiO_2$/SiC interface at the gate region of the MOSFETs are still far from the ideal behavior [2]. In particular, the most critical issues affecting the 4H-SiC MOSFETs performances are related to the low channel mobility and the threshold-voltage instability, mainly attributed to the high interface state density ($D_{it}$) near the conduction band edge ($E_C$) of 4H-SiC and to the presence of near interface oxide traps (NIOTs)[3,4]. The NIOTs are electrical active defects in the gate oxide close to the interface, which can trap carriers via tunneling mechanisms from the SiC substrate.

Using photo-stimulated tunnelling, Afanas'ev *et al.* [5,6] proposed a model of NIOTs at an almost identical energy position (2.77 eV below the $E_C$ of $SiO_2$) in different MOS structures both in Si and SiC. However, due to the fact that in 4H-SiC this oxide trap level is almost aligned to the bottom of the conduction band, the NIOTs can trap free carriers when electrons are attracted from the semiconductor substrate toward the insulator interface (as it occurs in the MOSFET channel). Thus, the NIOTs are often considered to be co-responsible, together with the interface states, for the low electron mobility in 4H-SiC MOSFETs[7]. However, the non-unique fabrication processing kept a lot of open question concerning the nature of the NIOTs in $SiO_2$/SiC materials. Hence, the location, distribution and chemical nature of the NIOTs have attracted the interest of the SiC scientific community and remains on of the hot topic for the materials science community.

Several alternative hypotheses to the Afanas'ev *et al.* have been proposed to explain the formation of the NIOTs and their nature, e.g., the presence of carbon clusters[8] and/or a carbon-rich transition layer (up to 10nm thick) at the $SiO_2$/SiC interface formed during thermal oxidation of SiC[9], the presence of oxygen vacancies[10] or nitrogen-related defects in $SiO_2$[11], etc. However, some earlier published results are not confirmed and unanimously accepted. Clearly, the chemical nature of the NIOTs in gate oxides on SiC remains a debated topic. In addition, their location inside the SiO$_2$ layer is another important information, since it determines their response time and, hence, has an impact on the MOSFETs channel conduction.

In this context, we have recently shown that the presence of negatively charged NIOTs in the SiO$_2$ layer can be responsible of the observed anomalous temperature dependence of the gate current in lateral MOSFETs[12]. In particular, introducing a modified Fowler Nordheim (FN) tunneling model to consider the change of the oxide electric field during neutralization of the NIOTs by holes injection enabled us to estimate a density of $N_{NIOT}=2\times10^{11}$ cm$^{-2}$ [12]. However, the nature and the position of the NIOTs with respect to the SiO$_2$/SiC interface have not been discussed in that work.

Recently, adapting the Yuan's distributed circuit model (which assumes a uniform distribution of NIOTs) [13], Zhang et al.[14] modified the model and assumed an exponentially decaying distribution of the NIOTs from the SiO$_2$/4H-SiC interface. In particular, from the fit of the frequency dependent capacitance and conductance measurements collected in MOS capacitors under strong accumulation, they estimated an amount of NIOTs of ~11×10$^{20}$cm$^{-3}$eV$^{-1}$ close to the SiC conduction band edge for the as-grown oxide[14]. The amount of NIOTs was reduced after a post-oxidation-annealing (POA) in NO down to ~3×10$^{20}$cm$^{-3}$eV$^{-1}$, which is typically adopted to increase the channel mobility. Furthermore, these authors also discussed the effect of the POA time[14]. In particular, the NIOTs distribution in the oxide is shrunk increasing the POA time from 10 min up to 1 hour. It decreases exponentially in both cases but it is reaching one tenth of the interface initial value at 2 nm or at 1 nm from the SiC interface as a function of the duration of the POA.

Recently, C. X. Zhang, et al.[15] used TCAD simulations to demonstrate that the low-frequency noise of the MOSFET is caused primarily by interface traps and DFT simulation to attribute these traps to carbon vacancy clusters and nitrogen dopant atoms at or near the SiC/SiO$_2$ interface; but in that paper a chemical direct investigation of the NIOTs is not reported yet. Moreover, M. S.

Dautrich, et al.,[16] employed a magnetic resonance technique to study the deep trapping states extending below the SiC/SiO$_2$ interface into the SiC bulk. However, the authors clearly mention that their (referred to the NIOTs) "*electronic properties are yet unknown*". This demonstrated that the topic remains still unclear and deserves further investigation combining electrical and chemical measurements with very high spatial resolution.

Finally, Fujino and Kita[17] presented a method to characterize the NIOTs in MOS capacitors using transient capacitance (C-t) measurements. In particular, the comparison between the measurements at room and low temperatures enabled them to determine the responses of the shallow and deep traps in the oxide.

In the present work, an electrical and chemical cross characterization of the NIOTs at the SiO$_2$/4H-SiC interface with a sub-nm resolution is presented. The electrical active defects within the gate oxide were investigated by carrying out transient capacitance (*C-t*) measurements, estimating the effective relaxation time of the NIOTs within the oxide. In addition, scanning transmission electron microscopy (STEM) and sub-nm resolution electron energy loss spectroscopy (EELS) probed the chemical composition of the SiO$_2$/4H-SiC near-interface region, associating the NIOTs to carbon-related defects in the first nanometer of a non-stoichiometric SiO$_x$ matrix in the *Non-Abrupt* (*NA*) SiO$_2$/4H-SiC interface.

### EXPERIMENTAL DETAILS

Lateral MOSFETs were fabricated on 4°-off-axis n-type (0001) 4H-SiC epitaxial layers with a doping concentration of $1\times10^{16}$ cm$^{-3}$ onto a heavy doped 4H-SiC substrates. The p-well consisted of an Al-implanted region with an acceptor level of $N_A\sim10^{17}$cm$^{-3}$ activated at 1650°C. The gate oxide was a 40 nm thick SiO$_2$ layer that was subjected to a post-deposition-annealing (PDA) treatment in N$_2$O-ambient at 1150°C for four hours to achieve acceptable values of the channel mobility[18].

The capacitance voltage (C-V) and the transient capacitance (C-t) characteristics of the devices were measured in a CASCADE Microtech probe station, using a Keysight B1505A parameter analyzer.

The lateral MOSFETs were used as electrical test vehicle to control at glance both the electron and the hole injection in the insulator and to perform the temperature dependent transient capacitance measurements (*C(t)*).

STEM analyses were performed in a state of the art (Cs)-probe corrected JEOL ARM200CF at a primary beam energy of 200 keV operating in scanning mode. Image are acquired in dark field z-contrast configuration using an annular dark-field detector. Electron Energy Loss (EELS) Spectroscopy Images are collected with a Gatan Quantum spectrometer in dual EELS configuration for energy drift correction. The energy dispersion was set to 0.25 eV/pixel in order to have all the three elements edge (100eV for silicon, 285eV for carbon and 530 eV for oxygen) in the same spectrum. The energy resolution – the half height width of the zero-loss peak – results in this configuration is 1.2 eV. The 3D spectrum image datacubes were acquired with a pixel size of 0.08 nm and pixel time of 0.02s in fast spectroscopy mode. Carbon and oxygen elemental maps are extracted using a power-law background subtraction and 30eV signal windows. TEM samples were prepared by mechanical polishing followed by low energy (0.5 keV) ion milling and 72 hours in a high temperature (150°C) degassing station under vacuum condition ($10^{-7}$ hPa). Finally, the lamella was treated for 60 s in plasma $O_2$ to remove the residual surface carbon contamination (see supplementary material S.1). The sample was prepared in the direction orthogonal to the surface steps related to the off-axis substrate in order to get flat and clean images of the interface.

**RESULTS AND DISCUSSION**

Firstly, C-V measurements have been performed on lateral MOSFETs operating in "gate-controlled-diode" (GCD) configuration[19]. In this operation mode, the source, drain, and body electrodes of the MOSFET are grounded, while the gate electrode becomes the anode of the GCD. In this way, the minority carriers (electrons) are supplied to the p-well from the n-type regions (source and drain). This method allows to provide minority carrier in the MOSFET body region keeping the semiconductor nearly at the equilibrium during the transient capacitance measurements.

Fig. 1 shows the C-V curves collected at 1kHz on the MOSFET in GCD configuration sweeping the gate bias from inversion (positive $V_G$) toward accumulation (negative $V_G$) and backward. In Fig. 1, a hysteresis between the forward and backward gate bias sweeps in a lateral MOSFET operated in GCD configuration is clearly visible. In previous works, it has been shown that the hysteresis between the forward and backward curves is partially ascribed to the presence of slow NIOTs in the system and fast interface states close to the valence band edge of the p-type SiC[20,21]. It can be argued that the system tries to reach the equilibrium, at each gate bias value of the sweep, in a certain time. Thus, as shown in Fig. 1, the hysteresis between the C-V curves tends to be eliminated and the variation is given by a transient capacitance at fixed $V_G$.

Fig. 2a shows the schematic procedure to collect the transient capacitance (*C-t*) measurement. The MOSFET is kept in inversion at $V_G$=+15V for 120s. Then, the gate is biased at the measurement value $V_{meas}$ (i.e., the flat band voltage $V_{FB}$= - 6V) for the measurement time. Thus, the *C(t)* curve, shown in Fig. 2b can be collected.

According to Fujino and Kita[17]– that have successfully employed the method on 4H-SiC MOS capacitors – the trapping/detrapping of the NIOTs produces a transient capacitance *C(t)* (collected in this work from t=0s and t=600s) that can be described by *extended-Debye relaxation model* [22] using the equation:

$$\Delta C \equiv |C(t) - C_{eq}| \cong |\Delta C_0| exp\left\{-\left(\frac{t}{\tau_{eff}}\right)^\beta\right\} \quad (1)$$

where *C(t)* is the capacitance at each time, $C_{eq}$ is the capacitance at the equilibrium (i.e, after 600s in our case), $\tau_{eff}$ is the characteristic tunneling time needed for the carrier transport back and forth from the NIOTs and the semiconductor and $0<\beta<1$ is a fitting parameter[17]. Indeed, the characteristic tunneling time is related to the spatial location of the NIOTs with respect to the $SiO_2$/SiC interface.

Eq. 1 can be rewritten in terms of $\Delta C/\Delta C_0$ in order to compare transient capacitance data saturating at different $C_{eq}$ values, as it occurs by changing the measurement temperature. Fig. 3 shows the experimental transient capacitance $\Delta C/\Delta C_0$ collected at different temperatures i.e. –30°C up to +100°C. The experimental data collected at each temperature can be fitted using Eq.1. The fits are also reported in Fig. 3 as dashed lines. Interestingly, the fitting of the experimental data resulted in the same characteristic tunnelling time $\tau_{eff} \approx 15s$ at all temperatures (from – 30°C up to +100°C). A constant characteristic time $\tau_{eff}$ indicates the occurrence of a temperature-independent relaxation mechanism, which in turn is the proof that a tunneling mechanism rules the discharge of the NIOTs. It has to be emphasized that the measured $\tau_{eff}$ represents the time needed to discharge the deepest NIOTs involved in the capacitance transient of an unknown distribution of defects within the $SiO_2$.

Fig. 4a shows the variation of the NIOTs relaxation time vs the gate bias at room temperature. This behaviour motivates why the C-t measurements at different temperature (from -30°C up to +100°C) were carried out at $V_G$=-6V (Fig.3). As shown in Fig. 4a, at $V_G$=-6V there is a minimum for the relaxation time ($\tau$=15s) that indicates the alignment between the NIOTs level and the Fermi level (see Fig. 4b). In fact, such alignment can minimize the tunnelling time for the electron from the NIOTs site to the semiconductor. Furthermore, Fig. 4b shows the MOS band diagram including the fast interface traps ($D_{it}$) and the slow NIOTs. Typically, the $D_{it}$ relaxation time is in the 10-100 μs range. Hence, the $D_{it}$ relaxation time is much shorter than those experimental relaxation times shown in Fig. 4a that consequently are associated to the NIOTs.

To get further insight into the chemical nature of the NIOTs responsible of the transient capacitance just shown before, sub-Ångström resolution STEM analyses combined with EELS have been used to monitor the chemical environment of the first nanometer of insulator from the 4H-SiC interface.

Figs.5a shows the high resolution cross-section dark field (DF) of the $SiO_2$/4H-SiC interface under investigation. Fig. 5b shows the spectrum image signal DF of the $SiO_2$/4H-SiC and the simultaneously collected elemental maps derived from EELS spectra. In particular, Fig. 5b shows the superimposed carbon, oxygen and nitrogen maps collected in the scanned region. The information on the composition can be interpreted averaging the signal on all the scanned lines. The results are depicted in Fig. 5c. The first step of our analysis was the precise identification of the $SiO_2$/4H-SiC interface. Looking at the DF signal – taken from Fig.5b – across the scanned region (dark line in Fig. 5c) 2.5 Å spaced periodic peaks can be observed. The position of the $SiO_2$/4H-SiC interface, i.e., where the SiC crystal starts to be oxidized, was assumed to be at the last (ideal) crystalline 4H-SiC plane. This position is labeled as the origin of the X-axis.

The chemical element profiling shows also the presence of nitrogen atoms segregated at the interface of the 4H-SiC and $SiO_2$. This result confirms the occurrence of nitrogen incorporation in the 4H-SiC during POA, which is known to induce a "counter doping" effect in the channel region[23,24]. On the other hand, a peak is visible in the DF signal profile on the $SiO_2$ side of the interface. This DF peak, having a lower intensity compared with those belonging to the 4H-SiC bulk substrate, represents the partially ordered silicon atoms of the insulating $SiO_2$ layer.

The above scenario was further confirmed looking at the EELS sub-stoichiometric $SiO_x$ map obtained using a 4 eV energy window between 99 and 103 eV (Fig. 5c). In fact, in this energy range only the silicon atoms not completely surrounded by oxygen can give a contribution to the EELS spectrum above the detection limit. On the other hand, purely oxidized $Si^{+4}$ oxidized atoms contribute to the EELS only at an energy larger than 104 eV. Here, it is possible to notice the progressive change in the $SiO_x$ and oxygen profiles across the $SiO_2$/4H-SiC interface, thus suggesting the presence of a *Non-Abrupt* (*NA*) $SiO_2$/4H-SiC interface with a non-ideal stoichiometry (sub-oxidized) composed of residual non-fullyoxidized SiC plane. As a consequence, the carbon profile shows a decreasing tail within the oxide that is wider than one nanometer. Following these experimental results, we can associate the NA interface to a partial substitution of carbon atoms both by oxygen and nitrogen atoms, leaving a complex mixture of Si, C, N and O species.

The resulting carbon profile (Fig. 5c ) – that is 2-3 SiC atomic planes wide – can be correlated to the electrical active defects (NIOTs) responsible of the transient capacitance discussed before. In particular, in the following section, the electrical behavior of the NIOTs and the chemical composition of the NA interface will be correlated.

Paulsen et al. [25] described the relaxation time $\tau(x)$ of traps located at a distance $x$ from the semiconductor interface using the following equations:

$$\tau(x) = \frac{m_1^* x \left(1 + \frac{1}{2\eta_1 x}\right)}{2\pi \eta_2 \hbar^3 D_{it}} exp(2\eta_1 x) \quad (2)$$

where

$$\eta_1^2 = \frac{2m_1^*(\phi_B - E_V + E_f)}{\hbar^2} \quad (3)$$

and

$$\eta_2^2 = \frac{2m_2^*(E_V + E_f)}{\hbar^2} \quad (4)$$

where $\phi_B$ is the energy discontinuity between the semiconductor and the oxide, $m_1^*$ and $m_2^*$ are the effective masses for electrons in the oxide and semiconductor (because at the flat band voltage the negatively charged NIOTs release electrons via-tunnelling toward the semiconductor) respectively, and $E_f$, is the position of the Fermi level in the semiconductor. Furthermore, $D_{it}$ is the interface state density close to the 4H-SiC valence band edge (for the details see the supplementary material S.2 referring to [19]), $\hbar$ is the reduced Plank constant. Using the values of $D_{it}$ reported in S.2 and the literature values of $m_1^*$ and $m_2^*$ [26], Eq. 2 can be solved as a function of the distance x between the NIOTs and the SiO$_2$/4H-SiC interface, as shown in Fig. 6.

Since we have early discussed that the NIOTs discharge occurs via-tunnelling mechanisms, those tunnelling events are independent no matter what is the position of the single trap with respect of the others. Thus, according to Eq. 2, the times needed to tunnel from a NIOTs at a certain distance $x_1$ are much smaller than the times needed to tunnel a distance $x_2$ if $x_1<x_2$. Hence, the experimentally measured effective relaxation time is due to the NIOTs deeper into the $SiO_2$ matrix. (see Fig. 5c). Given the characteristic time obtained from the transient capacitance measurements ($\tau_{eff} \approx 15s$) it is possible to estimate the maximum distance between NIOTs and $SiO_2$/4H-SiC interface involved in the experimental transient capacitance phenomena, namely, the thickness of the insulator responsible of the NIOTs discharge. According to Eq. 2, the $\tau(x)$ times can be calculated and depicted in Fig. 6. Considering the experimental value $\tau_{eff} \approx 15s$, a maximum tunneling distance from the 4H-SiC of 1.3 nm contributing to the capacitance transient ($\tau_{eff}$) can be determined (Fig. 6). This result corroborates our hypothesis, i.e., that the region from 0 nm till 1.3 nm within the $SiO_2$, depicted as NA interface in Fig. 5c, is the one containing the C-based defects responsible of the transient capacitance effects.

Fig. 7 shows the EELS spectra collected in three regions: at 3 nm far from the $SiO_2$/4H-SiC interface in the bulk of the $SiO_2$ (blue curve), at 3 nm far from the $SiO_2$/4H-SiC interface in the bulk of the 4H-SiC (red curve), and in the NA interface region. Obviously, no carbon is detected in the $SiO_2$ bulk (see also supplementary material S.1). On the other hand, the presence of a single peak at ~292 eV (i.e., in the σ* region) in the bulk of the 4H-SiC can be used as a marker of the $sp^3$ hybridization. In the TL the presence of a π* plateau (rather than a sharp π* peak) is highlighted in Fig. 7. The simultaneous presence of the σ* peak and of the π* plateau suggests that a mixed $sp^2/sp^3$ carbon hybridization is present in the NA interface transition region. Furthermore, the progressive variation of both the oxygen and silicon with different oxidation states within the NA interface region (Fig. 5c) are indicative of a complex scenario. In fact, according to qualitative EELS spectra simulations (see supplementary material S.3), a mixed $sp^2/sp^3$ carbon coordination with silicon, oxygen and nitrogen can explain the experimental results shown in Fig.7. In literature, single-carbon related oxide defects in the $SiO_2$/4H-SiC system have been debated[27,28]. More recently, ab-initio calculation also on carbon dimers C=C and also more complex defects (C-C=C) were presented[29]. However, further study on the mixed $sp^2$ and $sp^3$ carbon hybridizations coordinated with silicon, oxygen and nitrogen atoms are needed to understand the physics of the NIOTs and predict their role in the threshold voltage instabilities issue in 4H-SiC power MOSFETs.

**CONCLUSION**

In conclusion, in this paper the NIOTs in the $SiO_2$/4H-SiC system were investigated by carrying out transient temperature dependent capacitance transient (C-t) measurements to estimate the characteristic decay-time of the electrically charged traps. The NIOTs relaxation is ruled via tunneling mechanisms. Using a extended-Debye relaxation model allowed us to determine a maximum depth of 1.3nm for the NIOTs responsible for the transient phenomena. The $SiO_2$/4H-SiC interface has been chemically investigated employing EELS analyses. High resolution atomic profiling revealed a carbon containing non-stoichiometric $SiO_x$ layer, having thickness of about 1.3 nm in the non-abrupt interface region. The EELS C-K edge spectra collected in the not-abrupt interface pointed out the possibility to have carbon in a mixed $sp^2$/$sp^3$ hybridization realistically coordinated with silicon, oxygen and nitrogen atoms.

**Acknowledgments**

This work was carried out in the framework of the ECSEL JU project WInSiC4AP (Wide Band Gap Innovative SiC for Advanced Power), grant agreement n. 737483.

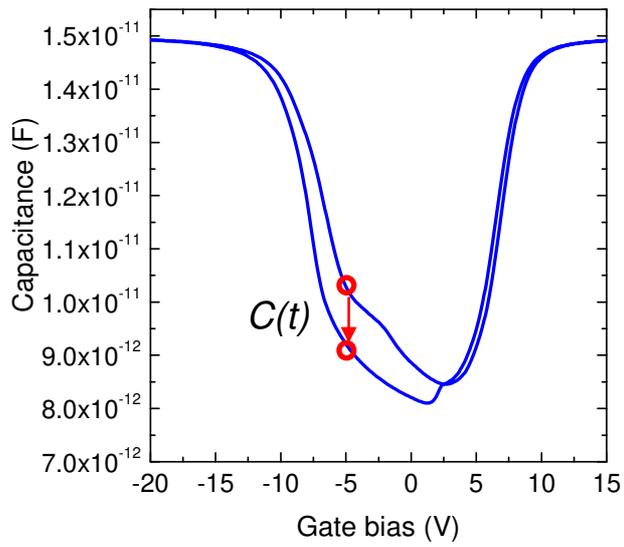

Figure 1 Inversion-to-accumulation and backward 1kHz C-Vs of MOSFETs in gate controlled diode configuration.

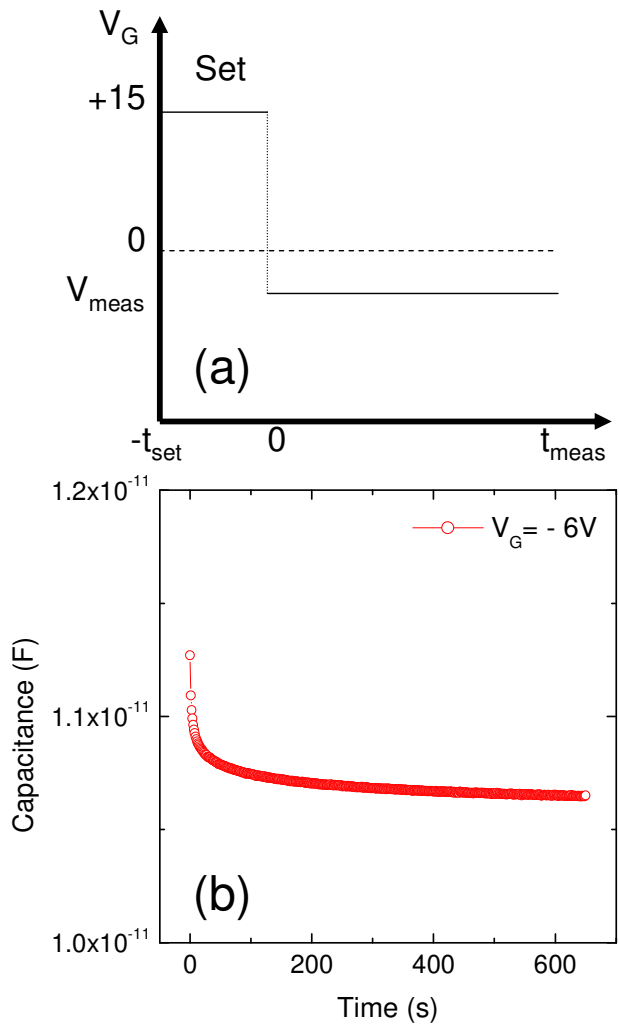

Figure 2 (a) Schematic of the experimental procedure to collect the transient capacitance (b) signal C(t).

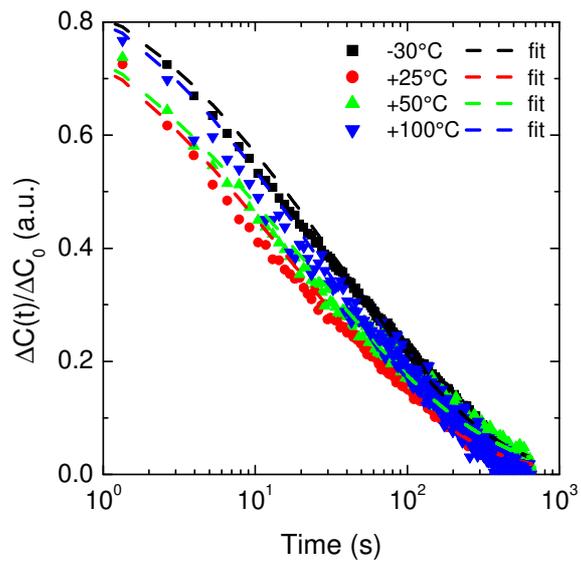

Figure 3 Experimental $\Delta C/\Delta C_0$ vs time at different temperatures. The fits of the data using Eq. 1 are also reported.

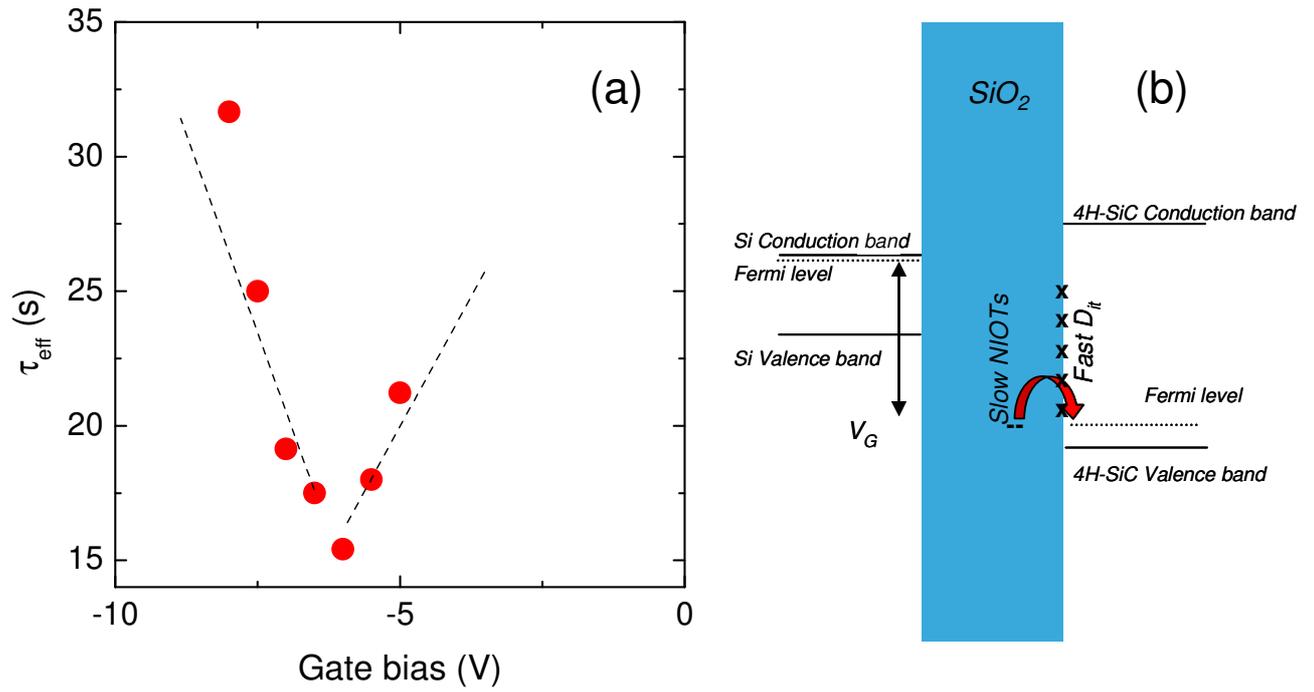

Figure 4. (a) Relaxation time vs gate bias measured at room temperature. (b) Schematic MOS band diagram including the presence of fast $D_{it}$ states and slow NIOTs states that resonate when are aligned to the Fermi level.

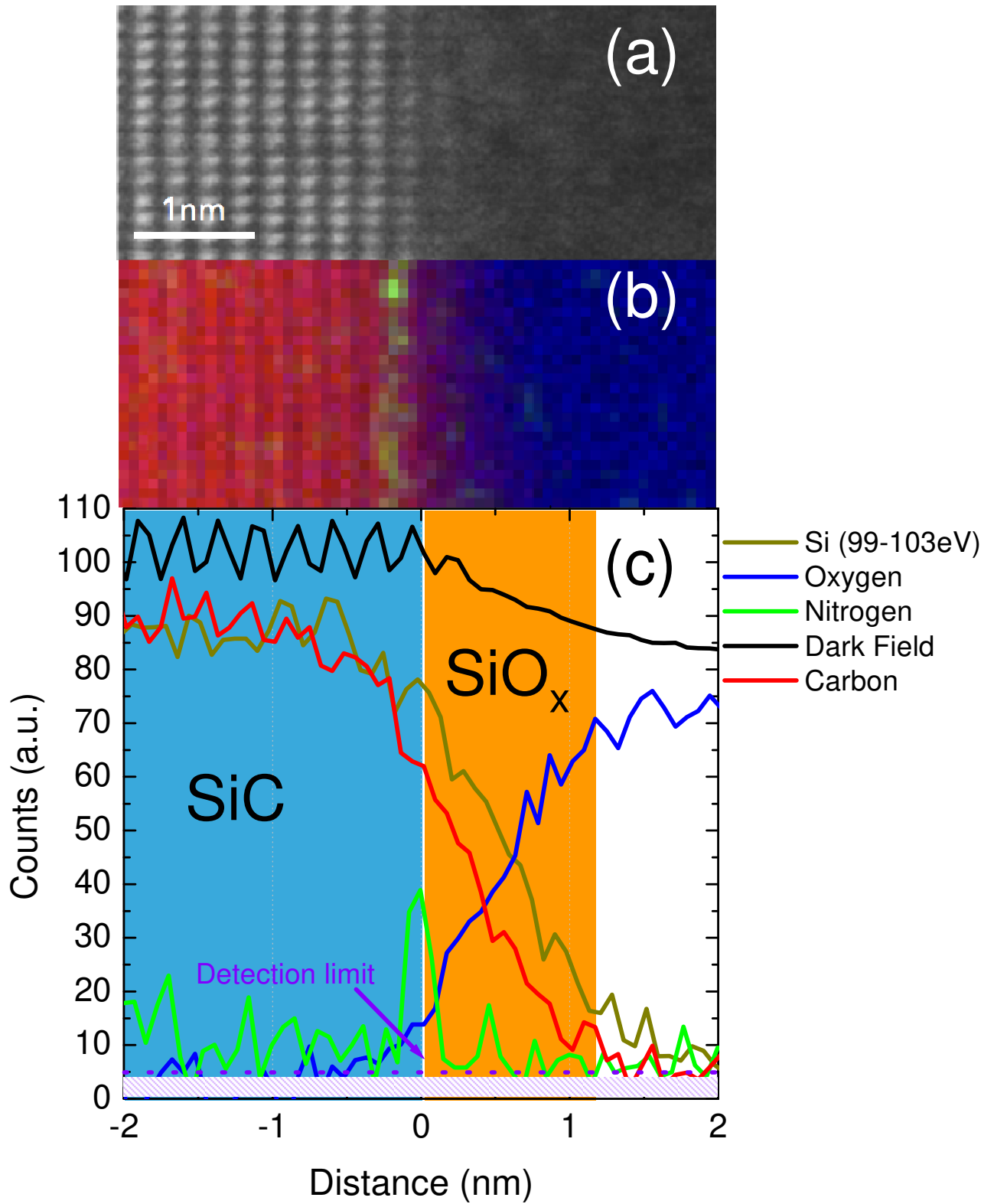

Figure 5. (a) High resolution dark field cross section STEM image across the SiO$_2$/4H-SiC interface. (b) Carbon, Oxygen and Nitrogen superimposed EELS maps. (c) Dark Field, Silicon, Carbon, Oxygen and Nitrogen EELS spectra averaged on all the scanned line; Bulk SiC is highlighted in light blue and the Not-Abrupt interface region in orange.

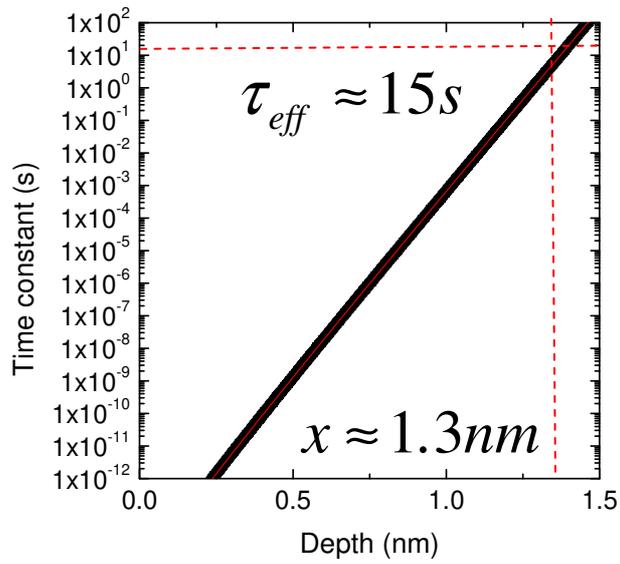

Figure 6: Tunnelling time constant as a function of the depth between the NIOTs and the $SiO_2$/4H-SiC interface, calculated using Eq. 2. The experimental value $\tau_{eff}$=15s correspond to a depth x=1.3nm.

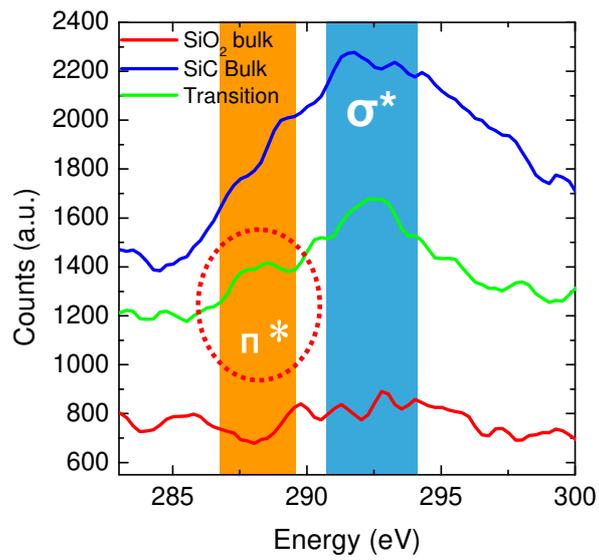

Figure 7. EELS Carbon spectra collected in the three different regions of the STEM lamella; in the SiC bulk (red), in the SiO$_2$ (blue) and in the not-abrupt interface transition region (green). Energetic regions related to σ* and π* bounds are highlighted.